\documentclass[final,1p,times]{elsarticle}

\usepackage{amssymb}

\journal{Nuclear Physics A}

\begin{document}

\begin{frontmatter}



\title{Pseudoscalar-Meson Octet-Baryon Coupling Constants \\
from two-flavor Lattice QCD}


\author[label1]{Toru T. Takahashi}
\author[label2]{G{\" u}ray Erkol}
\author[label3]{Makoto Oka}

\address[label1]{Yukawa Institute for Theoretical Physics, Kyoto University,
Sakyo, Kyoto 606-8502, Japan}
\address[label2]{Laboratory for Fundamental Research, Ozyegin University, Kusbakisi Caddesi No:2 Altunizade, \\ Uskudar Istanbul 34662 Turkey}
\address[label3]{Department of Physics, H-27, Tokyo Institute of Technology, Meguro, Tokyo 152-8551 Japan}

\begin{abstract}
We evaluate the pseudoscalar-meson octet-baryon coupling constants and the corresponding axial charges in eight channels ($\pi N\!N$, $\pi\Sigma\Sigma$, $\pi\Lambda\Sigma$, $K\Lambda N$, $K \Sigma N $, $\pi\Xi\Xi$, $K\Lambda\Xi$ and $K\Sigma\Xi$) in lattice QCD with two flavors of dynamical quarks. The parameter $\alpha\equiv F/(F+D)$ representing the SU(3)-flavor symmetry is computed at each u,d-quark hopping parameter and at the flavor-SU(3) symmetric point where the three quark flavors are degenerate at the physical $s$-quark mass. In particular, we obtain $\alpha=0.395(6)$ at the SU(3) symmetric limit. The quark-mass dependences of the coupling constants are obtained by changing the $u$- and the $d$-quark masses and we find that the SU(3)-flavor symmetry is broken by only a few percent at each quark-mass we employ. 

\end{abstract}


\end{frontmatter}



\section{Introduction}

It is long after the Quantum ChromoDynamics (QCD) was established as the underlying theory of the strong interaction in hadrons or nuclei. On the other hand, QCD-based description of hadron physics has not yet been successful due to the strong coupling nature of QCD. The low-energy QCD dynamics is far from the perturbative regime, and nonperturbative analyses are essentially needed. Among potentially workable methods, lattice QCD has been established as a powerful first-principle method for QCD. Lattice QCD has been well developed and almost at the mature level. Nowadays, the physical point can be directly accessed with lattice QCD, and it can cast light on hadrons or hadronic interactions directly from the viewpoint of QCD.

In hadron physics, meson-baryon coupling constants are one of the most important ingredients. Baryon are the building blocks of our world, and meson-baryon coupling constants provide a measure of baryon-baryon interactions in terms of One Boson Exchange (OBE) models. In phenomenological potential models, the meson-baryon coupling constants are determined so as to reproduce the nucleon-nucleon, hyperon-nucleon and the hyperon-hyperon interactions in terms of, {\it e.g.}, OBE models. It is then an important issue to determine such coupling constants directly from QCD, the underlying theory of the strong interactions. Lattice QCD is the only method that provides a first-principle calculation which can be directly compared with experiments, and it serves as a valuable tool to determine the hadron couplings in a model-independent way.


Meson-baryon coupling constants directly reflect the flavor structure in hadron sectors. In the SU(3)-flavor~[SU(3)$_F$] symmetric limit, one can classify the pseudoscalar-meson--octet-baryon coupling constants in terms of two (undetermined) parameters: the $\pi N\!N$ coupling constant and the $\alpha=F/(F+D)$ ratio~\cite{deSwart:1963gc};
\begin{eqnarray}
g_{\pi N\!N}=g,\quad g_{\pi\Sigma\Sigma}=2 g \alpha,
\quad
g_{\pi\Lambda\Sigma}= \frac{2}{\sqrt{3}}g(1-\alpha),
\quad
g_{K\Lambda N}=-\frac{1}{\sqrt{3}}g(1+2\alpha),
\nonumber
\\
g_{K\Sigma N}=g(1-2\alpha),
\quad
g_{\pi\Xi\Xi}=-g(1-2\alpha),
\quad
g_{K\Lambda\Xi}=\frac{1}{\sqrt{3}}g(4\alpha-1),
\quad
g_{K\Sigma\Xi}=-g.
\label{su3rel}
\end{eqnarray}
This systematic classification governs all the meson-baryon couplings in the case of SU(3)$_F$ symmetric limit. However as we move from the symmetric case to the realistic one, SU(3)$_F$ breaking inevitably occurs as a result of the mass difference between $u,d$- and $s$-quarks. The broken symmetry no longer guarantees a pattern for the meson-baryon coupling constants, and the breaking pattern can be determined only from the underlying QCD dynamics. They therefore should be individually calculated directly based on QCD. In this work we extract the coupling constants $g_{\pi N\!N}$, $g_{\pi\Sigma\Sigma}$, $g_{\pi\Lambda\Sigma}$, $g_{K\Lambda N }$, $g_{K\Sigma N}$, $g_{\pi\Xi\Xi}$, $g_{K\Lambda\Xi}$ and $g_{K\Sigma\Xi}$ (denoted by $g_{MBB^\prime}$ hereafter) by means of lattice QCD with two flavors of dynamical quarks. 

\section{Meson-Baryon coupling constants and axial charges}

The pseudoscalar current matrix element is written as
\begin{equation}
	\langle {\cal B}({\bf p}) |P(0)|{\cal B}^\prime({\bf p}^\prime) \rangle = g_P(q^2)\bar{u}({\bf p}) i \gamma_5 u({\bf p}^\prime),
\end{equation}
where $g_P(q^2)$ is the pseudoscalar form factor, $q_\mu=p_\mu^\prime-p_\mu$ is the transferred four-momentum and $P(x)=\bar{\psi}(x)i\gamma_5 \frac{\tau_3}{2}\psi(x)$ is the pseudoscalar current. We compute this matrix element using the standard techniques on the lattice~\cite{Alexandrou:2006mc,Alexandrou:2007zz,Yamazaki:2008py,Lin:2007ap,Sasaki:2008ha,Erkol:2008yj,Erkol:2009ev}.
The baryon interpolating fields employed in this study are given as
\begin{small}
\begin{eqnarray}
&&\eta_N(x)=\epsilon^{abc}[u^{T a}(x) C \gamma_5 d^b(x)]u^c(x),
\quad
\eta_\Sigma(x)=\epsilon^{abc}[s^{T a}(x) C \gamma_5 u^b(x)]u^c(x), \\
&&\eta_\Lambda(x)=\frac{1}{\sqrt{6}}\epsilon^{abc}
\left\{[u^{T a}(x) C \gamma_5 s^b(x)]d^c(x)-[d^{T a}(x)C\gamma_5 s^b(x)]u^c(x)
+2[u^{T a}(x) C \gamma_5 d^b(x)]s^c(x)\right\}, \\
&&\eta_\Xi(x)=\epsilon^{abc}[s^{T a}(x) C \gamma_5 d^b(x)]s^c(x),
\end{eqnarray}
\end{small}
where $C=\gamma_4\gamma_2$ and $a$, $b$, $c$ are the color indices. 
Calculating $g_P^L(q^2)$ ($g_P$ on the lattice) at the momentum transfers ${\bf q}^2 a^2=n (2\pi/L)^2$ (for the lowest nine $n$ points), where $L$ is the spatial extent of the lattice, we obtain the meson-baryon form factor $g_{M B B^\prime}(q^2)$ via the relation,
\begin{equation}
	g_P^L(q^2)=\frac{G_M\,g_{M B B^\prime}(q^2)}{m_M^2-q^2},
\label{MBff}
\end{equation}
assuming that the pseudoscalar form factors are dominated by the pseudoscalar-meson poles. Here $G_M\equiv\langle {\rm vac} | P(0) | M \rangle$ is extracted from the two-point mesonic correlator $\langle P(x)P(0)\rangle$. Finally we extract the meson-baryon coupling constants $g_{M B B^\prime}=g_{M B B^\prime}(0)$ by means of a monopole form factor:
\begin{equation}
	g_{M B B^\prime}(q^2)=g_{M B B^\prime} \frac{\Lambda^2_{M B B^\prime}}{\Lambda^2_{M B B^\prime}-q^2}.	
\end{equation}

We employ a $16^3\times 32$ lattice with two flavors of dynamical quarks and use the gauge configurations generated by the CP-PACS collaboration~\cite{AliKhan:2001tx} with the renormalization group improved gauge action and the mean-field improved clover quark action. We use the gauge configurations at $\beta=1.95$ with the clover coefficient $c_{SW}=1.530$, which give a lattice spacing of $a=0.1555(17)$ fm ($a^{-1}=1.267$ GeV), which is determined from the $\rho$-meson mass. The simulations are carried out with four different hopping parameters for the sea and the $u$,$d$ valence quarks, $\kappa_{sea},\kappa_{val}^{u,d}=$ 0.1375, 0.1390, 0.1400 and 0.1410, which correspond to quark masses of $\sim$ 150, 100, 65, and 35~MeV, and we use 490, 680, 680 and 490 such gauge configurations, respectively. The hopping parameter for the $s$ valence quark is fixed to $\kappa_{val}^{s}=0.1393$ so that the Kaon mass is reproduced~\cite{AliKhan:2001tx}, which corresponds to a quark mass of $\sim 90$~MeV. We employ smeared source and smeared sink, which are separated by 8 lattice units in the temporal direction. Source and sink operators are smeared in a gauge-invariant manner with the root mean square radius of 0.6 fm. All the statistical errors are estimated via the jackknife analysis.

\begin{figure}[th]
	\includegraphics[width=0.5\textwidth]{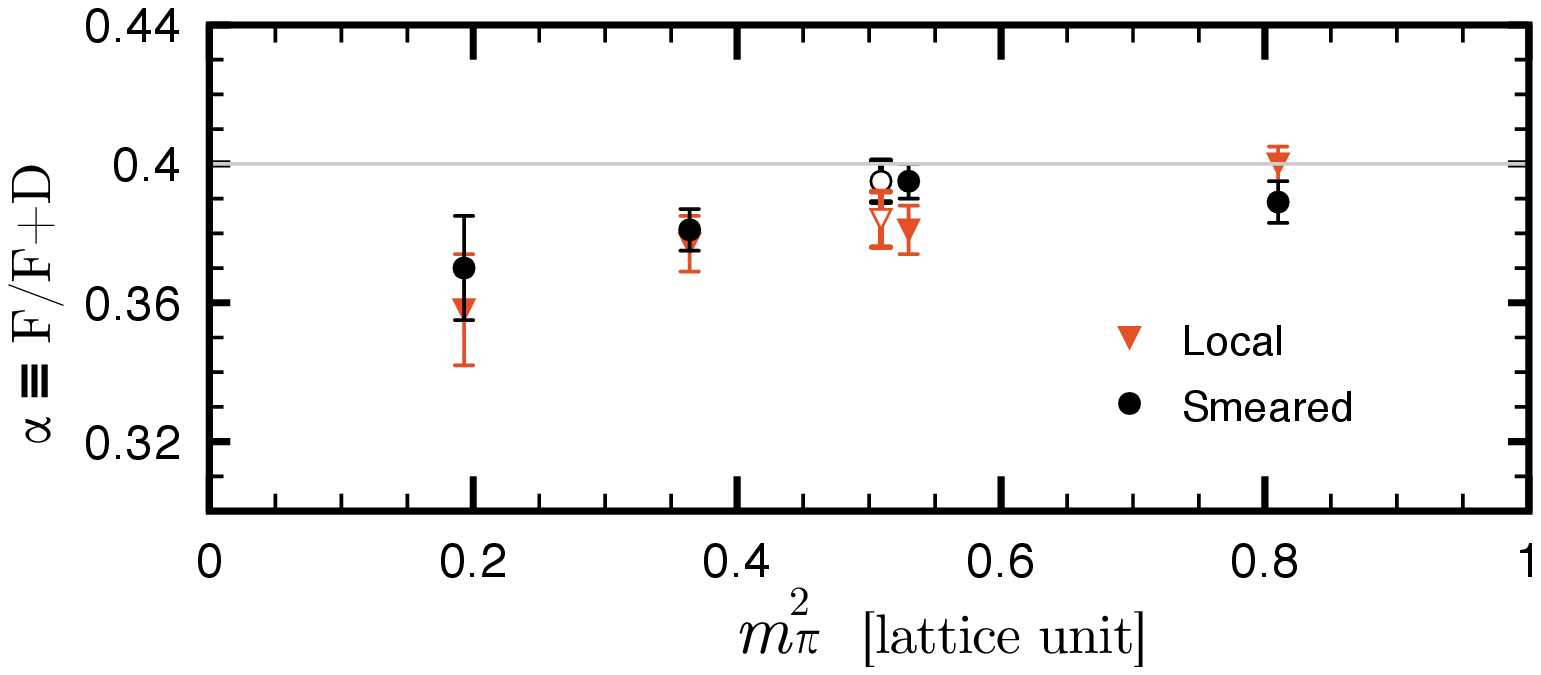}
	\includegraphics[width=0.5\textwidth]{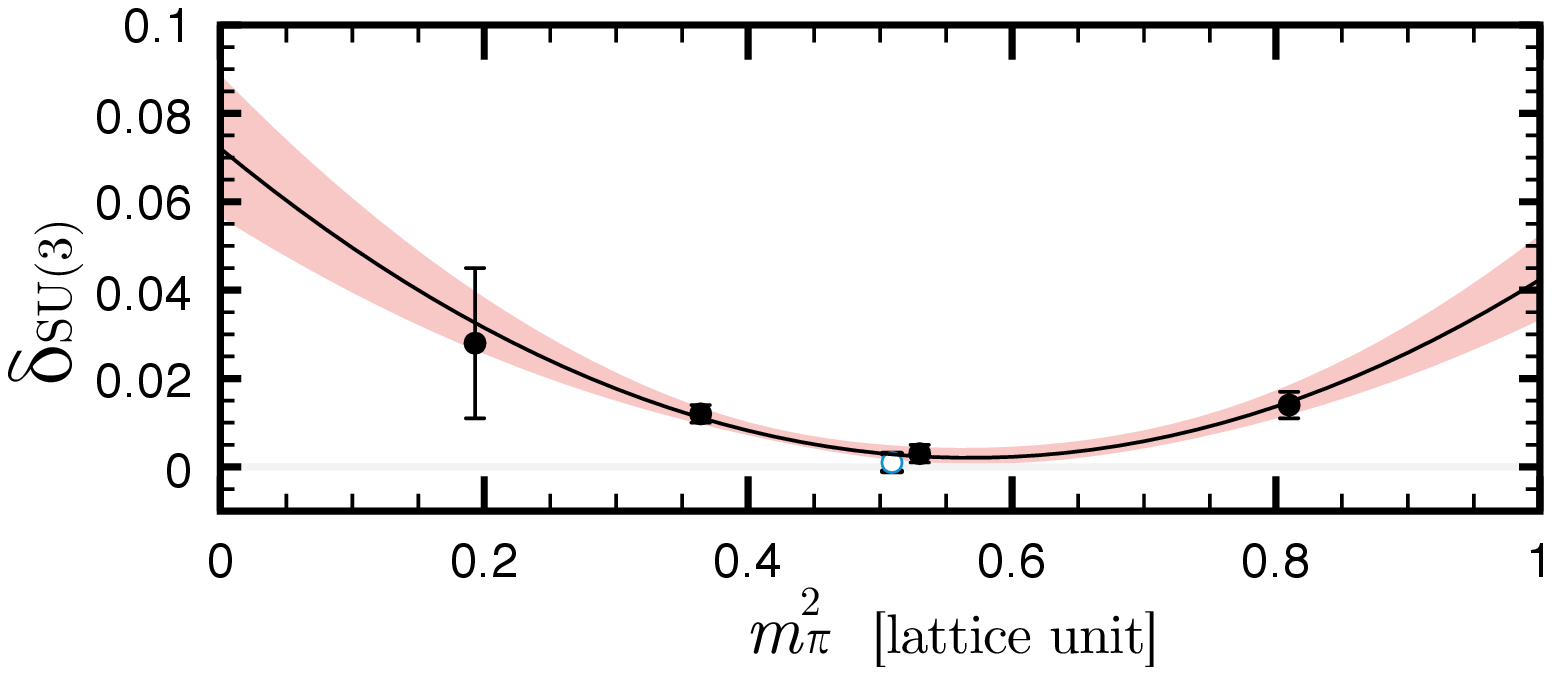}
	\caption{\label{alpha_MB} {\bf LEFT: }$\alpha_A=F/F+D$ ratio extracted by a global fit of eight meson-baryon coupling constants is plotted as a function of $m_\pi^2$. {\bf RIGHT: }The SU(3)-breaking parameter~\cite{Erkol:2008yj,Erkol:2009ev} extracted by a global fit of eight meson-baryon couplings. The empty circles in both figures denote the data at the SU(3)$_F$ symmetric limit.}
\end{figure}	

In the SU(3)-flavor symmetric case, where $\kappa_{val}^{u,d}\equiv\kappa_{val}^s = 0.1393$ and the SU(3) relations are exact, all the couplings are well reproduced with $\alpha=0.395(6)$, which is obtained by a global fit of the lattice data using Eq.(\ref{su3rel}). The obtained value of $\alpha$ is consistent with that in the SU(6) spin-flavor symmetry ($\alpha=2/5$)~\cite{Pais:1966}, which is the symmetry based on the nonrelativistic quark model. As we move to the chiral limit, the mass difference between $u,d-$ and $s$-quarks gets large, and the fitted $\alpha$ gradually decreases toward the chiral limit. The quark-mass dependence of $\alpha$ can be found in the left panel in Fig.~\ref{alpha_MB}. We here define and evaluate the SU(3) breaking parameter $\delta_{\rm SU(3)}$, which represents to what extent the SU(3) relations are broken. (See Refs.\cite{Erkol:2008yj,Erkol:2009ev} for the detail.) The breaking parameter $\delta_{\rm SU(3)}$ is plotted in the right panel in Fig.~\ref{alpha_MB}. Interestingly enough, $\delta_{\rm SU(3)}$ remains {\cal O}(1)\% in the quark-mass range we consider, which suggests for the pseudoscalar-meson couplings of the octet baryons that SU(3)$_F$ is a good symmetry. We try a quadratic fit of $\delta_{{\rm SU}(3)}$ and find that $\delta_{{\rm SU}(3)}=0.072(16)$ in the chiral limit. The SU(3)$_F$ breaking effect seems to appear in $\alpha$ rather than in SU(3)$_F$ relations.

\begin{figure}[th]
	\includegraphics[width=0.5\textwidth]{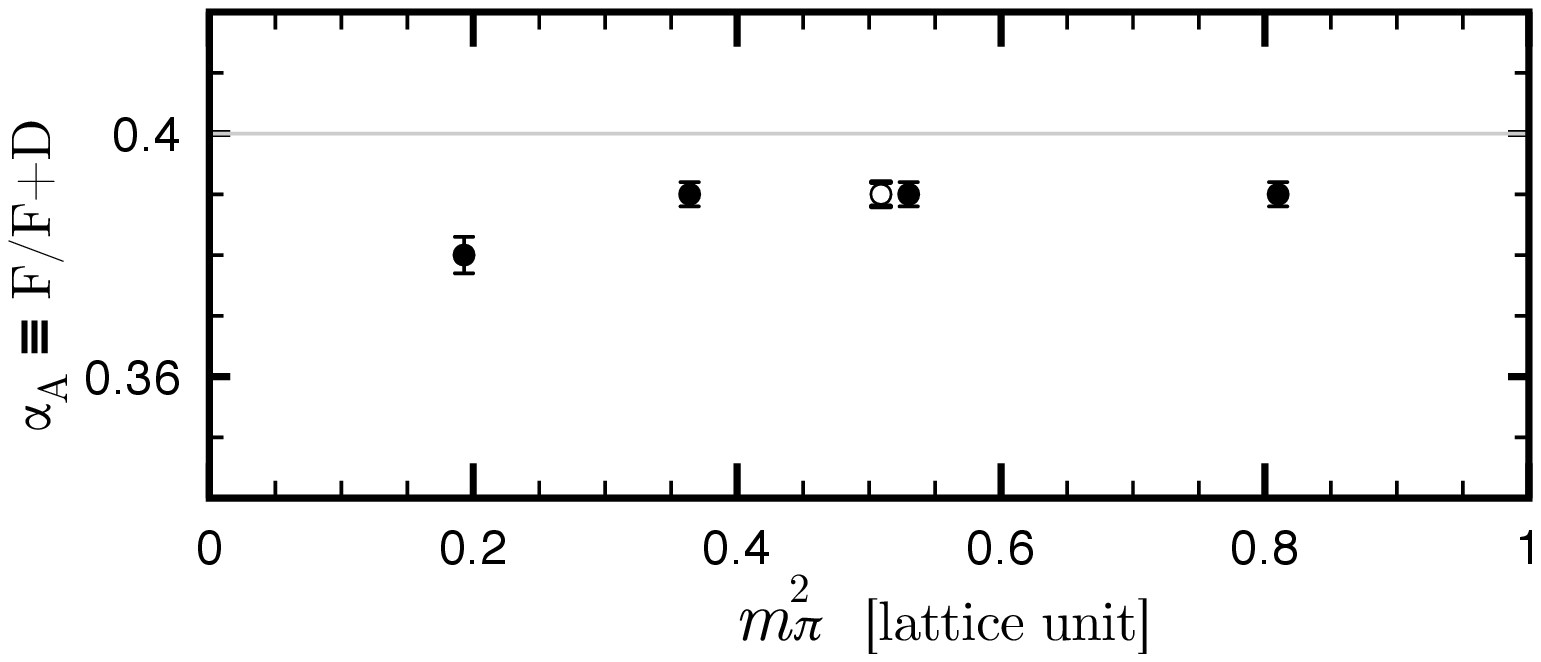}
	\includegraphics[width=0.5\textwidth]{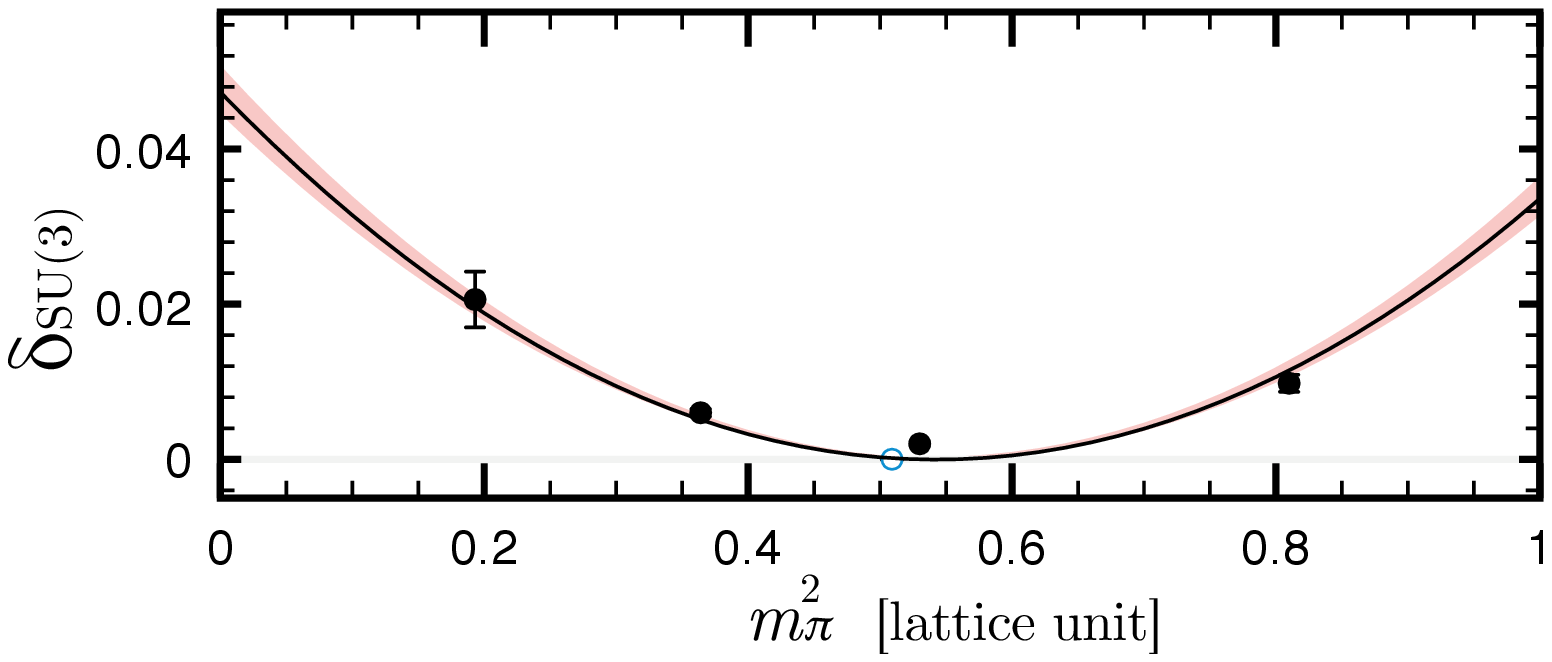}
	\caption{\label{alpha_GA} {\bf LEFT: }$\alpha_A=F/F+D$ ratio extracted by a global fit of eight axial charges is plotted as a function of $m_\pi^2$. {\bf RIGHT: }The SU(3)-breaking parameter~\cite{Erkol:2008yj,Erkol:2009ev} extracted by a global fit of eight axial charges. The empty circles in both figures denote the data at the SU(3)$_F$ symmetric limit.}
\end{figure}	
We also extract the corresponding eight axial charges
($g_{A,NN}$, $g_{A,\Sigma\Sigma}$, $g_{A,\Lambda \Sigma}$, $g_{A,\Lambda N}$, $g_{A,\Sigma N}$, $g_{A,\Xi\Xi}$, $g_{A,\Lambda\Xi}$ and $g_{A,\Sigma\Xi}$),
and repeat our analysis for them.
Fig.~\ref{alpha_GA} shows $\alpha$ and $\delta_{\rm SU(3)}$ extracted by global fits of the eight axial charges. The SU(3) parameter $\alpha$ is found to be 0.390(2) in the SU(3) limit, and we observe essentially the same tendency as meson-baryon couplings.

\section{Summary}
We have evaluated the pseudoscalar-meson--octet-baryon coupling constants as well as the corresponding axial charges in eight channels ($\pi N\!N$, $\pi\Sigma\Sigma$, $\pi\Lambda\Sigma$, $K\Lambda N$, $K \Sigma N $, $\pi\Xi\Xi$, $K\Lambda\Xi$ and $K\Sigma\Xi$), in two-flavor lattice QCD with the hopping parameters $\kappa_{sea},\kappa_{val}^{u,d}=$ 0.1375, 0.1390, 0.1400 and 0.1410, which correspond to quark masses of $\sim$ 150, 100, 65, and 35~MeV. Our results indicate, for both of pseudoscalar-meson couplings and axial charges of the octet baryons, 
\begin{enumerate}
\item SU(3)$_F$ parameter $\alpha\equiv F/(F+D)$ is 0.395(6) for the meson-baryon couplings (0.390(2) for the axial charges) in the SU(3)$_F$ symmetric limit, which is consistent with the prediction from SU(6) spin-flavor symmetry ($\alpha = 2/5$). 
\item SU(3)$_F$ relations hold well, which is broken by only a few percent (at least) in the 35 MeV to 150 MeV range of the light quark masses. SU(3)$_F$ is hence a good symmetry in the quark-mass range we consider.
\item The parameter $\alpha$ gradually decreases towards the chiral limit, and SU(3)$_F$-breaking effect appear in $\alpha$ rather than SU(3)$_F$ relations themselves.
\end{enumerate}

\section*{acknowledgment}
All the numerical calculations were performed on NEC SX-8R at CMC, Osaka university, on SX-8 at YITP, Kyoto University, on BlueGene at KEK, and on TSUBAME at TITech. The unquenched gauge configurations employed in our analysis were all generated by CP-PACS collaboration~\cite{AliKhan:2001tx}. This work was supported in part by the Yukawa International Program for Quark-Hadron Sciences (YIPQS), by the Japanese Society for the Promotion of Science under contract number P-06327 and by KAKENHI (17070002, 19540275, 20028006 and 21740181).






\end{document}